\documentclass[aip,apl,reprint]{revtex4-2}

\usepackage{graphicx}
\usepackage{amsmath}
\usepackage{bm}
\usepackage{lipsum}
\usepackage[mathlines,columnwise]{lineno}
\begin{document}
%\runninglinenumbers

\title{Self-selected phase-matched second harmonic generation in nonlinear optical materials: from phenomenon to applications}

\author{Niklas Dömer}
\affiliation{Institute of Physics, Department of Mathematics/Informatics/Physics, Osnabrück University, 49076 Osnabrück, Germany}

\author{Tobias Hehemann}
\affiliation{Institute of Physics, Department of Mathematics/Informatics/Physics, Osnabrück University, 49076 Osnabrück, Germany}

\author{Felix Sauerwein}
\affiliation{Institute of Physics, Department of Mathematics/Informatics/Physics, Osnabrück University, 49076 Osnabrück, Germany}

\author{Sebastian Inckemann}
\affiliation{Leibniz-Institut für Kristallzüchtung (IKZ), 12489 Berlin, Germany}

\author{Steffen Ganschow}
\affiliation{Leibniz-Institut für Kristallzüchtung (IKZ), 12489 Berlin, Germany}

\author{Mirco Imlau$^\ast$}
\affiliation{Institute of Physics, Department of Mathematics/Informatics/Physics, Osnabrück University, 49076 Osnabrück, Germany}

\begin{abstract}

Self-selected phase-matched second harmonic generation is introduced as an all-optical probe of refractive-index dispersion in birefringent nonlinear optical materials. Rather than requiring wavelength or angular tuning, the exposure with a spectrally broad, intense ultrashort pulse allows the material to self-select the fundamental spectral component that satisfies the type-I noncritical phase-matching condition. This produces a narrow peak in the second harmonic spectrum whose position is governed by the refractive indices and is therefore highly sensitive to material parameters that affect the optical dispersion. We demonstrate the application of this phenomenon for the optical inspection of stoichiometry and temperature gradients in technologically relevant lithium niobate, as well as composition inhomogeneities in newly grown lithium niobate–tantalate solid solutions. These results establish self-selected phase-matched second harmonic generation as a rapid, non-contact method for inspecting nonlinear optical materials, with potential relevance for bulk crystals, wafers, and thin-film platforms.

\end{abstract}

\maketitle

\begin{flushleft}
\textbf{Author emails:}
\mbox{ndoemer@uni-osnabrueck.de};
\mbox{tobias.hehemann@uni-osnabrueck.de};
\mbox{felix.sauerwein@uni-osnabrueck.de};
\mbox{sebastian.inckemann@ikz-berlin.de};
\mbox{steffen.ganschow@ikz-berlin.de};
\mbox{mimlau@uni-osnabrueck.de}.
\end{flushleft}

\noindent $^\ast$Corresponding author: Mirco Imlau, Institute of Physics, 
Department of Mathematics/Informatics/Physics, Osnabr\"uck University, 
49076 Osnabr\"uck, Germany. Email: mimlau@uni-osnabrueck.de; 
Tel.: +49 541 / 969 2654.

\section{Introduction}

Phase-matched second harmonic generation (SHG) provides a direct optical probe of refractive-index dispersion and is therefore intrinsically sensitive to material parameters such as composition, stoichiometry, temperature, and defect-related changes  \cite{ horst_determination_1998}. In noncritical phase-matched SHG (NCPM SHG), the phase-matching condition is fulfilled along a principal crystallographic axis, such that the wave-vector mismatch between the fundamental and second harmonic waves vanishes without angular tuning \cite{hobden_phasematched_1967}. For uniaxial crystals, this corresponds to conditions of the form $n_o(2\lambda)=n_e(\lambda)$ for negatively birefringent materials, and $n_o(\lambda)=n_e(2\lambda)$ for positively birefringent materials, considering the ordinary ($n_o$) and extraordinary ($n_e$) refractive indices, respectively. Here, $\lambda$ denotes the SHG wavelength, such that the fundamental wavelength is $2\lambda$.

In conventional narrowband NCPM SHG, the phase-matching condition is usually accessed by tuning the temperature \cite{362711_temp_phasematching_lbo, Jundt_Temp_shg, biaggio_refractive_1992_temp_tuning_shg}. Multiple studies have shown that, for narrowband excitation, the phase-matching temperature depends on the crystal stoichiometry and can vary within a single crystal \cite{bergman_curie_1968, fay_dependence_1968, schmidt_composition_1991}. In this work, we transfer this established sensitivity from the temperature-tuning domain to the spectral domain: broadband femtosecond excitation provides a continuum of fundamental wavelengths, from which the material selects the spectral component that fulfills the NCPM condition. As a result, a pronounced peak appears in the SHG spectrum at the phase-matched wavelength. The position of this peak is determined by the refractive-index dispersion and can therefore serve as an all-optical spectral marker for material variations, provided that the corresponding dispersion relation is known or calibrated.

We demonstrate this concept using lithium niobate (LN) and lithium niobate--tantalate (LNT) as model systems. LN is one of the most established and technologically important nonlinear and electro-optic materials, serving as a central platform for bulk nonlinear optics, electro-optic devices, integrated photonics and quantum optics \cite{weis_1985, doi:10.1126/science.abj4396}. Together with related solid solutions such as LNT, it provides a particularly relevant model system for developing optical probes of refractive-index dispersion \cite{Bashir10092023, bollmers_surface-near_2024}. Their optical properties are highly sensitive to stoichiometry, composition and temperature, all of which directly affect the refractive indices and their dispersion \cite{Wood_2008, PhysRevB.87.195208,cryst10110973,cryst16010001}. Reliable characterization of such variations is therefore essential, particularly for mixed crystals and spatially inhomogeneous samples. A range of optical and analytical techniques has been developed for this purpose, each offering specific advantages but also limitations \cite{wohlecke_optical_1996, noncolinear_SHG_paper_betzler}.

\begin{figure*}[!htb]
\includegraphics[width=\textwidth]{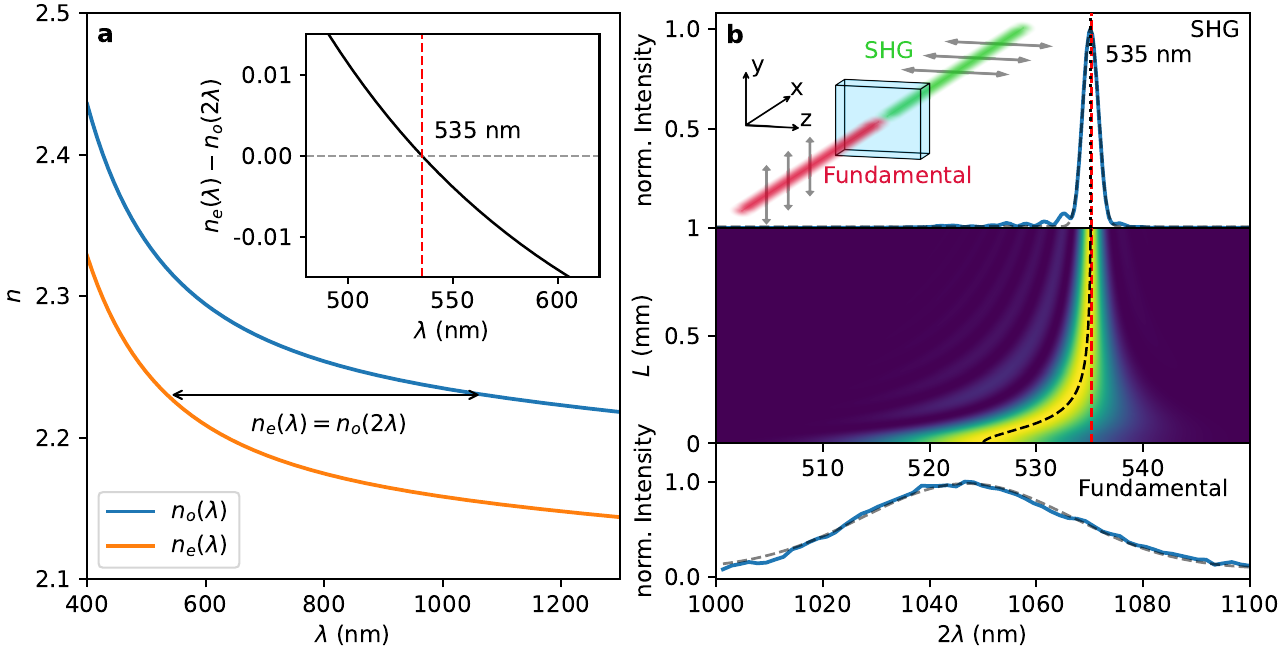}
\caption{a: Ordinary ($n_o$) and extraordinary ($n_e$) refractive indices of congruent lithium niobate calculated from Sellmeier equations \cite{PhysRevB.48.15613}. The NCPM condition is fulfilled when $n_e(\lambda)-n_o(2\lambda)=0$, as highlighted in the inset, showing a zero crossing at $\sim 535\,\mathrm{nm}$. 
b: Broadband excitation and self-selected NCPM SHG response. The measured fundamental spectrum (bottom, FWHM $\sim 46\,\mathrm{nm}$) produces a narrow measured SHG peak (top, FWHM $\sim 1.7\,\mathrm{nm}$) at the NCPM wavelength. Gray dashed lines indicate Gaussian fits. The middle panel shows the calculated normalized SHG response according to Eq.~(1) for the measured fundamental spectrum propagating through a $1\,\mathrm{mm}$ thick crystal; each calculated spectrum is normalized individually. The red dashed vertical line marks the calculated NCPM wavelength from a, while the black dashed curve shows the peak position obtained from Gaussian fits to the calculated spectra as a function of interaction length. The vertical dotted line in the measured SHG spectrum marks the fitted Gaussian center. The inset sketches the used collinear type-I geometry.}
\label{fig:shg_spectrum}
\end{figure*}

Lithium niobate is particularly well suited for demonstrating broadband self-selected NCPM SHG because its NCPM wavelength lies in the visible spectral range and produces a narrow, pronounced SHG feature. We first demonstrate the emergence of this spectral marker in congruent LN and compare the measured peak wavelength with Sellmeier-based calculations. We then validate the approach by temperature-dependent measurements on congruent and near-stoichiometric LN, before applying it to LNT mixed crystals with growth-induced compositional gradients and comparing spatial SHG peak-wavelength maps with $\mu$-XRF measurements. To assess the broader applicability of the concept, we further identify NCPM conditions in selected birefringent nonlinear crystals and summarize calculated and measured peak wavelengths.

The underlying concept is illustrated in Fig.~\ref{fig:shg_spectrum}. The ordinary $n_o$ and extraordinary $n_e$ refractive indices are shown as functions of wavelength based on Sellmeier equations for room-temperature congruent lithium niobate in Fig.~\ref{fig:shg_spectrum}a \cite{PhysRevB.48.15613}. NCPM occurs at wavelengths where the condition $n_e(\lambda)-n_o(2\lambda)=0$ is satisfied \cite{boyd_nonlinear_2008}. The inset highlights this condition, showing a zero crossing of $n_e(\lambda)-n_o(2\lambda)$ at $\sim 535\,\mathrm{nm}$.

The SHG spectral intensity can be described within the plane-wave and undepleted-pump approximation by
\begin{equation}
I_{\mathrm{SHG}}(\lambda) \propto d_{\mathrm{eff}}^2 I_{\mathrm{fund}}^2(2\lambda) L^2
\,\mathrm{sinc}^2\!\left(\frac{\Delta k(\lambda)L}{2}\right),  \label{eqn:SHG}
\end{equation}
where \(d_{\mathrm{eff}}\) is the effective nonlinear coefficient, $I_{\mathrm{fund}}(2 \lambda) $ the fundamental intensity, \(L\) the sample thickness, and \(\Delta k\) the phase mismatch \cite{boyd_nonlinear_2008}. For the present type-I \(o+o\rightarrow e\) phase-matching geometry, \(d_{\mathrm{eff}} = d_{31}\).
The inset in Fig. \ref{fig:shg_spectrum}b illustrates the corresponding beam geometry: the light propagates along the crystallographic $x$ axis of the $x$-cut crystal, with the fundamental field polarized along $y$ and the generated second harmonic field polarized along $z$, as indicated by the arrows on the beams.
Using the SHG wavelength \(\lambda\), the phase mismatch is given by
\begin{equation}
\Delta k(\lambda)=k_{\mathrm{SHG}}-2k_{\mathrm{fund}}
=\frac{2\pi}{\lambda}\left[n_e(\lambda)-n_o(2\lambda)\right]. \label{eqn:deltaK}
\end{equation}
Hence, the phase-matching condition is fulfilled for \(\Delta k(\lambda)=0\), i.e. when $n_e(\lambda)=n_o(2\lambda)$.

\begin{figure*}[!htb]
\includegraphics[width=\textwidth]{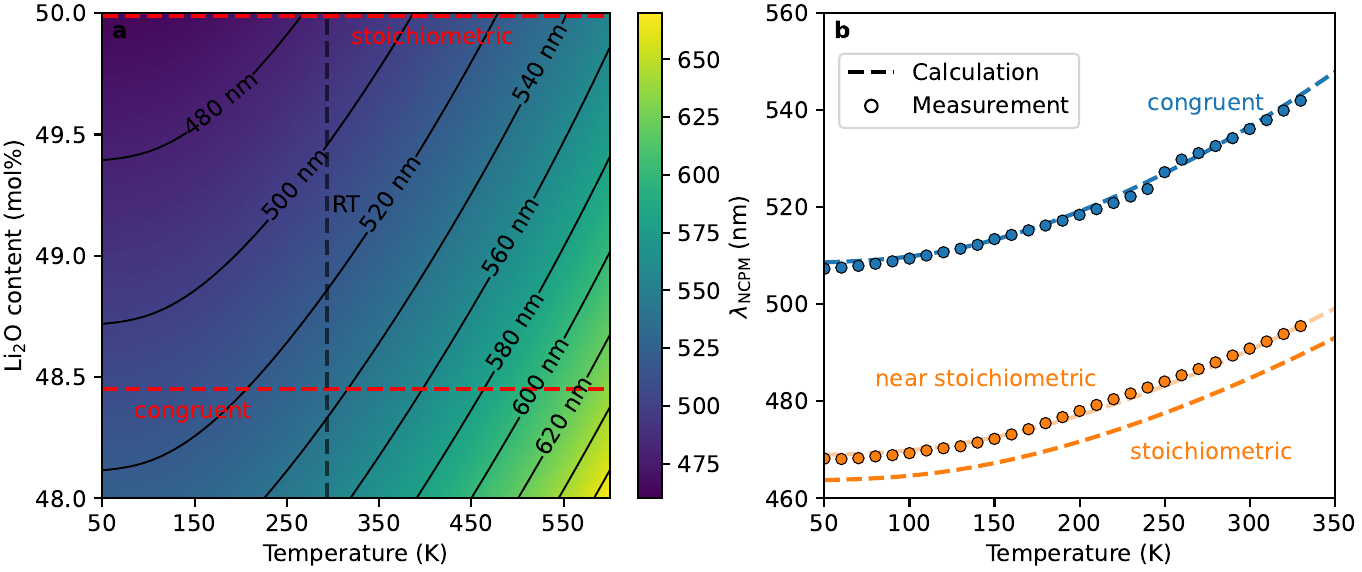}%{fig_temperature_li_map_and_measurements_kelvin.png}
\caption{a: Calculated resonance wavelength $\lambda_{\mathrm{NCPM}}$ as a function of temperature and lithium content in lithium niobate. The map is obtained from the condition $n_e(\lambda) - n_o(2\lambda) = 0$, with contour lines indicating constant resonance wavelengths. Horizontal dashed lines mark the lithium content corresponding to congruent  (48.45\,mol\%) and stoichiometric (50\,mol\%) material, while the vertical dashed line indicates room temperature (20$^\circ$C).
b: Temperature dependence of the NCPM wavelength for congruent and stoichiometric lithium niobate. Dashed lines represent calculated values at 50\,mol\% (stoichiometric), 48.45\,mol\% (congruent) and 49.8\,mol\% (near stoichiometric), while markers denote experimentally extracted peak positions from Gaussian fits to the SHG spectra.} 
\label{fig:compUndTemp}
\end{figure*}

\section{Results}

\subsection{SHG Spectra and Phase Matching}

The key experimental signature of broadband NCPM SHG is the emergence of a narrow SHG peak selected from the much broader fundamental spectrum. Figure~\ref{fig:shg_spectrum}b (lower panel) shows the fundamental spectrum of the OPA, with a full width at half maximum (FWHM) of $\sim 46\,\mathrm{nm}$ centered at $\sim 1050\,\mathrm{nm}$. In contrast, the corresponding second harmonic spectrum is significantly narrower, exhibiting a FWHM of $1.7\,\mathrm{nm}$ centered at $535\,\mathrm{nm}$. Notably, the SHG peak is not located at half the fundamental peak wavelength, reflecting the noncritical phase-matching condition at ambient temperatures. In addition, the fundamental and second harmonic beams propagate collinearly, indicating the absence of spatial walk-off. The collinear geometry, the absence of intentional angular tuning, and the spectral displacement of the SHG maximum from half the fundamental peak wavelength together identify the observed signal as NCPM SHG under ambient conditions.

Gaussian fits are used as a phenomenological and reproducible way to extract the peak position and spectral width (gray dashed lines). Although the ideal phase-matching response follows a sinc$^2$ dependence (Eq.~\ref{eqn:SHG}), the measured spectrum results from this response being weighted by the spectral envelope of the broadband fundamental. This produces an approximately Gaussian main peak, while the weak asymmetric spectral fringes originate from residual sinc$^2$ side lobes typically known from rocking curves. The measured SHG peak position, marked with a black dotted line, is in very good agreement with the calculated phase-matching wavelength.

To illustrate the origin of this spectral narrowing, Fig. \ref{fig:shg_spectrum}b (mid panel) shows the calculated SHG response obtained from Eq. (\ref{eqn:SHG}) as a function of SHG wavelength and interaction length. For very short interaction lengths, the generated spectrum is mainly governed by the spectral envelope of the broadband fundamental pulse. With increasing interaction length, the $\mathrm{sinc}^2$ phase-matching term increasingly suppresses non-phase-matched contributions, leading to a narrow spectral maximum close to the condition $\Delta k = 0$. 
The peak position extracted from Gaussian fits to the calculated spectra is shown as the black dashed line and rapidly converges to the calculated NCPM wavelength indicated by the red dashed line. In this case, already for an interaction length of $0.5\,\mathrm{mm}$, the deviation is below $0.2\,\mathrm{nm}$, demonstrating that $\lambda_{\mathrm{NCPM}}$ provides a robust spectral observable for the phase-matching condition.

\subsection{Dependence on Stoichiometry and Temperature}

In 1993, Schlarb and Betzler published a generalized Sellmeier equation for lithium niobate that simultaneously accounts for both lithium content and temperature \cite{PhysRevB.48.15613}. This model allows calculating the ordinary and extraordinary refractive indices, $n_o(\lambda)$ and $n_e(\lambda)$, for temperatures between 50\,K and 600\,K and for varying lithium concentrations. Based on these relations, the NCPM wavelength $\lambda_{\mathrm{NCPM}}$ can be determined as a function of both temperature and composition.

\begin{figure}[!b]
    \centering
    \includegraphics[width=\columnwidth]{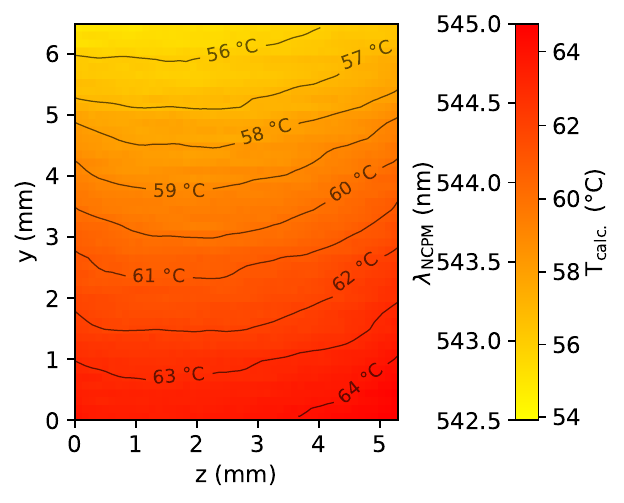}
    \caption{Spatial map of the NCPM wavelength $\lambda_{\mathrm{NCPM}}$ obtained from Gaussian fits to the measured spectra of a bottom-heated congruent lithium niobate crystal. The secondary colorbar and contour lines show the equivalent temperature $T_{\mathrm{calc.}}$ calculated from the Schlarb--Betzler Sellmeier model for $c_{\mathrm{Li}}=48.45\,\mathrm{mol}\%$ Li$_2$O.\label{fig:TemperatureMap}}
\end{figure}

The resulting dependence is shown in Fig.~\ref{fig:compUndTemp}a as a two-dimensional color-coded map. The horizontal axis represents the temperature (50--600\,K), while the vertical axis shows the $\mathrm{Li_2O}$ content in mol\% from 48 - 50\,mol\%. Red dashed lines indicate the stoichiometric composition ($\mathrm{Li_2O}=50\,\mathrm{mol}\%$) and the  congruent composition ($\mathrm{Li_2O} \approx 48.45\,\mathrm{mol}\%$\cite{Bryan_congruent}), noting that slightly different values for the latter are reported in the literature \cite{Sanna_surfaces, jundt_axial_2008, ZANATTA2022105736}.

The NCPM wavelength $\lambda_{\mathrm{NCPM}}$ is encoded in color, and contour lines are added to highlight its variation. Notably, neither horizontal nor vertical contour lines are observed, indicating that $\lambda_{\mathrm{NCPM}}$ depends nontrivially on both temperature and lithium content. Consequently, a single measurement of $\lambda_{\mathrm{NCPM}}$ cannot independently determine both quantities; however, if either temperature or lithium content is known, the other can be inferred.

For validation, we measured the NCPM wavelength $\lambda_{\mathrm{NCPM}}$ as a function of temperature between 50\,K and 330\,K for both a congruent and a stoichiometric lithium niobate crystal. The results are shown in Fig.~\ref{fig:compUndTemp}b, where the experimental data (circles) are compared to the calculated dependencies (dashed lines). 

\begin{figure*}[t]
\includegraphics[width=\textwidth]{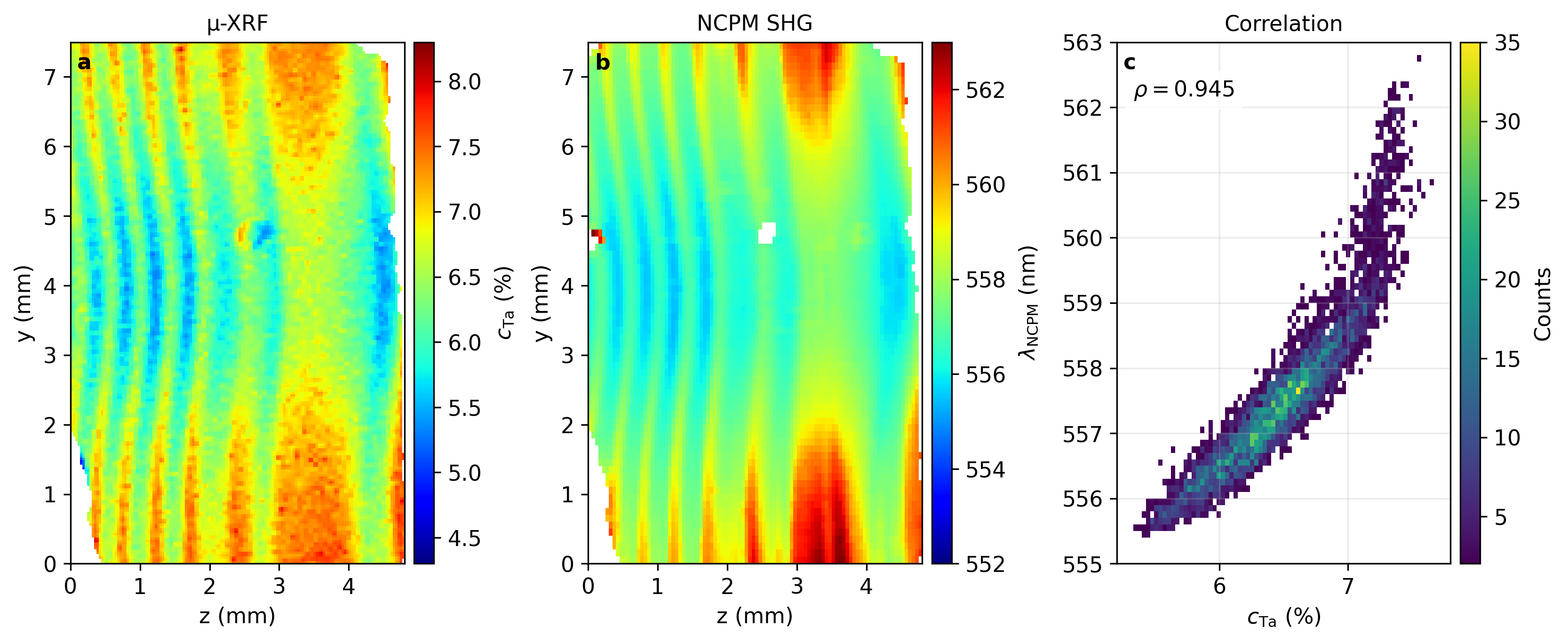}
\caption{a: $\mu$-XRF map of the tantalum concentration expressed as Ta/(Ta+Nb) for an LNT crystal, showing compositional variations between approximately $5$ and $8\,\mathrm{mol}\%$. 
b: Spatial map of the SHG peak wavelength $\lambda_{\mathrm{NCPM}}$ obtained from Gaussian fits to the measured spectra. White regions indicate positions with insufficient signal. 
c: Point-density plot of $\lambda_{\mathrm{NCPM}}$ versus tantalum concentration from the aligned $\mu$-XRF data. For each SHG pixel, the tantalum concentration was obtained by averaging all XRF pixels whose centers fall within the corresponding SHG pixel. The color scale represents the number of points per bin starting at 2, showing a clear, albeit non-linear, monotonic dependence.}
\label{fig:XRF_Map}
\end{figure*}

Excellent agreement is observed for the congruent crystal over the entire measured temperature range. In contrast, the measurements for the nominally stoichiometric sample exhibit a nearly constant offset with respect to the theoretical prediction. This deviation can be consistently explained by assuming a slightly lithium-deficient composition of $\sim 49.8\,\mathrm{mol}\%$, corresponding to a typical near-stoichiometric crystal \cite{lengyel_growth_2015}. Under this assumption, the measured values align well with the expected temperature dependence.

To demonstrate that the NCPM wavelength can also be used as a spatially resolved temperature observable, we imposed a spatial temperature gradient across a congruent lithium niobate crystal. The crystal was mounted in a clamp holder on a double-stack Peltier element and heated from below, such that the lower part of the crystal was warmer than the upper part. The heated sample was then scanned spatially, with each local SHG spectrum fitted to extract $\lambda_{\mathrm{NCPM}}$. Since room-temperature reference measurements revealed no detectable composition-induced variation of $\lambda_{\mathrm{NCPM}}$, spatial changes in $\lambda_{\mathrm{NCPM}}$ can be attributed to temperature and converted into a corresponding temperature $T_{\mathrm{calc.}}$ using the calibration shown in Fig.~\ref{fig:compUndTemp}. Figure~\ref{fig:TemperatureMap} shows the resulting spatial map, revealing a clear temperature gradient from bottom to top.

The slightly curved, parabola-like isotherms indicate additional lateral heating from the heat-conducting clamps, which weakly heat the crystal from the sides. Thus, the measured NCPM wavelength reproduces the expected two-dimensional temperature map and demonstrates the capability of NCPM SHG for contact-free thermal mapping.

\subsection{Spatial Mapping of Composition}

A recent study has shown that in lithium niobate-tantalate (LNT) mixed crystals, the extraordinary refractive index changes only weakly as the composition changes, while the ordinary refractive index increases with rising tantalum content \cite{hehemann_disperion_2026}. It is therefore expected that the NCPM wavelength shifts toward longer wavelengths as the tantalum content increases and - based on the available dispersion data - disappears at a nearly equimolar composition\cite{koppitz_second_2025}.

Consequently, spatial variations in tantalum concentration are expected to result in spatially varying phase-matching wavelengths. To investigate this, we study an x-cut LNT crystal with an average tantalum content of $\sim 6\,\mathrm{mol}\%$ (Ta/(Ta+Nb)), exhibiting pronounced growth striations.

A $\mu$-XRF map of the investigated region is shown in Fig.~\ref{fig:XRF_Map}a, revealing local composition variations in the range of approximately $5$--$8\,\mathrm{mol}\%$. The crystallographic $z$-axis is oriented horizontally. Front- and back-side $\mu$-XRF measurements showed no significant differences within the investigated region, suggesting that variations along the sample thickness are small compared with the lateral gradients.

The same region was investigated by self-selected NCPM SHG. At each position, the SHG spectrum was recorded and fitted with a single Gaussian in the region of interest. The extracted peak wavelength $\lambda_{\mathrm{NCPM}}$ was then mapped spatially, as shown in Fig.~\ref{fig:XRF_Map}b. White regions correspond to positions with insufficient signal, including sample edges and areas with optical damage.

To align the XRF and SHG wavelength maps, translation and rotation were optimized automatically by maximizing the Spearman correlation coefficient within the common sample area. For the correlation analysis, the XRF data were mapped onto the SHG grid by assigning to each SHG pixel the mean tantalum concentration of all XRF pixels whose centers fall within that SHG pixel. Regions without reliable SHG signal, such as sample edges or optically damaged areas, were excluded from the correlation analysis. A direct comparison reveals that the structural features and striations observed in the $\mu$-XRF data are clearly reproduced in the SHG peak wavelength map.

This correspondence is further quantified in Fig.~\ref{fig:XRF_Map}c, which shows the correlation between $\lambda_{\mathrm{NCPM}}$ and the tantalum concentration obtained from $\mu$-XRF as a point-density plot. The color scale represents the number of SHG pixels falling into a given interval of $\lambda_{\mathrm{NCPM}}$ and Ta concentration. A clear, albeit non-linear, correlation is observed, as reflected by the Spearman rank correlation coefficient $\rho = 0.945$. The observed scatter may arise from the higher spatial resolution of the XRF measurement, experimental noise, or intrinsic differences in probing depth, as SHG is sensitive to the bulk response whereas XRF is predominantly surface-sensitive. The overall agreement demonstrates that self-selected NCPM SHG can map compositional variations in LNT through their effect on the refractive-index dispersion.

\section{Discussion}

These results establish self-selected NCPM SHG as a robust probe of the NCPM condition and, therefore, as a probe of changes in refractive-index dispersion. The method does not measure stoichiometry, composition, temperature, or dopant concentration directly; rather, these quantities are inferred from $\lambda_{\mathrm{NCPM}}$ when their influence on the ordinary and extraordinary refractive-index dispersion is known from a model or established through calibration. This distinction is important for extending the approach beyond the congruent and near-stoichiometric LN crystals studied above.

Within lithium niobate, this extension is particularly relevant for doped crystals, where dopants can modify $n_o$, $n_e$, or their dispersion in different ways. This suggests that NCPM SHG may also be used to probe dopant-induced refractive-index variations or dopant inhomogeneities  \cite{cryst10110973, mizuno_temperature-dependent_2015}. As an experimental indication, we observe a pronounced NCPM SHG peak in Mg-doped stoichiometric LN, with a measured peak wavelength of $497\,\mathrm{nm}$, in good agreement with the calculated value of $494\,\mathrm{nm}$ \cite{Cleo_sln_Shoji:11} listed in Table \ref{tab:otherCrystals}. This shows that the spectral-marker concept is not restricted to undoped LN, although quantitative dopant mapping would require a calibrated relation between dopant concentration and refractive-index dispersion.

\begin{table}[tb]
\caption{
Calculated and measured SHG peak wavelengths for selected nonlinear optical materials.
Multiple rows for one material indicate different possible noncritical phase-matching
conditions. Corresponding measurements are shown in the supplementary section. \label{tab:otherCrystals}
}
\label{tab:otherCrystaly}
\begin{ruledtabular}
\begin{tabular*}{\columnwidth}{@{\extracolsep{\fill}}lp{0.34\columnwidth}cc}
Material &
Matching indices &
$\lambda_{\mathrm{NCPM}}^{\mathrm{calc.}}$ (nm) &
$\lambda_{\mathrm{NCPM}}^{\mathrm{meas.}}$ (nm) \\
\colrule

Mg:sLN\makebox[0pt][l]{\,\cite{Cleo_sln_Shoji:11}}  &
$n_x(\lambda)=n_z(2\lambda)$ &
$494$&
$497$ \\[0.4em]

LBO\makebox[0pt][l]{\,\cite{LBO_refractive_index}} &
$n_x(\lambda)=n_z(2\lambda)$ &
$276$  &
$276$ \\

&
$n_x(\lambda)=n_y(2\lambda)$ &
$343$ &
$-$ \\[0.4em]

KNbO$_3$\makebox[0pt][l]{\,\cite{KNB_refractive_index_Zysset:92}} &
$n_{\alpha}(\lambda)=n_\beta(2 \lambda)$ &
$491$&  
$491$ \\

&
$n_{\alpha}(\lambda)=n_{\gamma}(2 \lambda)$ &
$429$ &
$-$ \\[0.4em]

KTP\makebox[0pt][l]{\,\cite{KTP_Kato:02}} &
$n_y(\lambda)=n_z(2\lambda)$ &
$398$ &
$402$ \\

&
$n_x(\lambda)=n_z(2\lambda)$ &
$369$ &
$-$ \\
\end{tabular*}
\end{ruledtabular}
\end{table}

The applicability of the technique is governed by the available phase-matching geometry, the effective nonlinearity, birefringence, dispersion, transparency, and interaction length. A suitable crystal cut and polarization configuration must provide access to an index crossing with nonzero $d_{\mathrm{eff}}$, while the broadband fundamental spectrum must overlap the corresponding phase-matching wavelength. Moreover, the interaction length has to be sufficiently large for the phase-matching term in Eq.~(1) to spectrally filter the broadband response; in very thin samples, the SHG spectrum would be dominated by the fundamental envelope and the NCPM peak would become broader and less distinct as illustrated in Fig.\ref{fig:shg_spectrum}. If the birefringence is too large, the NCPM wavelength may be shifted into the ultraviolet, where linear absorption can suppress the SHG signal. Conversely, for weakly birefringent materials, such as lithium tantalate or tantalum-rich LNT, the phase-matching condition may no longer be fulfilled. In addition, reduced dispersion toward the infrared leads to a shallower zero crossing of the NCPM condition, resulting in broader and less pronounced SHG peaks according to Eq.~(2), which complicates reliable detection. The favorable situation in lithium niobate, where the NCPM wavelength lies in the visible spectral range and exhibits a pronounced spectral feature, therefore represents a particularly advantageous case.

To illustrate the broader applicability of the concept, Table \ref{tab:otherCrystals} summarizes calculated and measured NCPM peak wavelengths for selected birefringent nonlinear crystals. The table is not intended to be exhaustive, but shows that suitable index crossings occur in several technologically relevant materials. For biaxial crystals such as LBO, KNbO$_3$, and KTP, the three principal refractive indices provide several possible combinations of fundamental and second harmonic polarizations, and therefore multiple possible NCPM conditions within a single material. The Sellmeier calculations used to identify these crossings, together with representative measured SHG spectra, are shown in the Supplementary Material. Dashes in the measured-wavelength column indicate that no corresponding measurement was performed because a crystal with a suitable cut was not available in this study, not that the respective NCPM condition is experimentally inaccessible.

The same self-selection concept is not limited to noncritical phase matching. For a fixed critically phase-matched geometry, broadband excitation does likewise generate a spectral maximum at the wavelength satisfying $\Delta k(\lambda,\theta,\phi)=0$. This would broaden the applicability of the method, but at the cost of a more complex interpretation: the involved refractive indices become angle-dependent effective indices of the polarization eigenmodes rather than simple principal refractive indices.

In future work, this approach may be extended to microscopy-based implementations, enabling high-resolution all-optical mapping of material properties. Conventional confocal SHG microscopy usually relies on intensity-based contrast and does not explicitly exploit the spectral position of the SHG signal \cite{koppitz_second_2025}. This, however, has been shown to carry additional information on phase-matching conditions and material dispersion \cite{hegarty_tuning_2022}.

\section{Materials and methods}

The experiments were performed using a regeneratively amplified Ti:Sapphire laser system (Astrella, Coherent Corp., Saxonburg, PA, USA; repetition rate $1\,\mathrm{kHz}$, pulse duration $\tau_{\mathrm{p}} \approx 40\,\mathrm{fs}$) pumping an optical parametric amplifier (OPA, TOPAS Prime, Light Conversion, Vilnius, Lithuania), which provided broadband near-infrared excitation. Residual short-wavelength components from the OPA were suppressed using a long-pass filter (RG850, SCHOTT AG, Mainz, Germany). The beam was weakly focused onto the sample using a lens with a focal length of $f = 400\,\mathrm{mm}$, resulting in a beam diameter of approximately $200\,\mu\mathrm{m}$ FWHM at the crystal. For spatially resolved measurements, the sample was mounted on a two-dimensional translation stage. The average power was attenuated with neutral-density filters to below $2\,\mathrm{mW}$. The OPA center wavelength was adjusted for each sample to ensure spectral overlap with the expected NCPM condition. During the temperature-dependent measurements shown in Fig.~\ref{fig:compUndTemp}b, the fundamental spectrum was monitored and, if necessary, readjusted to maintain sufficient overlap with the NCPM wavelength. The fundamental polarization was oriented perpendicular to the crystallographic $z$-axis ($\vec{E} \perp \vec{z}$). Temperature-dependent measurements were performed in a closed-cycle helium cryostat (RW-3 with LTC 60 controller, Leybold GmbH, Cologne, Germany) equipped with plane-parallel optical windows.

The transmitted light was analyzed using spectrometers covering the UV--VIS (IsoPlane with PIXIS 2k and i3-060-300-P grating (600 G/mm), Princeton Instruments, Trenton, NJ, USA) and NIR (NIRQuest, Ocean Optics, Orlando, FL, USA) spectral ranges. Integration times were on the order of $1\,\mathrm{s}$ and were adjusted depending on the sample and signal strength. The SHG spectra were fitted with Gaussian functions in the spectral region of interest to extract the peak wavelength $\lambda_{\mathrm{NCPM}}$.

$\mu$-XRF measurements were performed using an M4 TORNADO system (Bruker Nano GmbH, Berlin, Germany). Elemental maps of Ta and Nb were used to determine the local tantalum fraction, $c_{\mathrm{Ta}} = \mathrm{Ta}/(\mathrm{Ta}+\mathrm{Nb})$. For comparison with the SHG measurements, the $\mu$-XRF maps were aligned to the SHG peak-wavelength maps by applying translations and rotations and optimizing the spatial correlation within the common sample area. The XRF data were then mapped onto the SHG grid by averaging all XRF pixels whose centers fall within the corresponding SHG pixel.

The investigated crystals include congruent lithium niobate (x-cut, thickness $1\,\mathrm{mm}$, SurfaceNet GmbH, Rheine, Germany), nominally stoichiometric lithium niobate (x-cut, thickness $0.5\,\mathrm{mm}$, SurfaceNet GmbH), and lithium niobate--tantalate which was grown using the Czochralski method at the Leibniz Institute for Crystal Growth (IKZ) (x-cut, thickness $0.92\,\mathrm{mm}$), as described in Bashir \textit{et al.}  \cite{Bashir10092023}. Additional crystals used for the supplementary measurements together with respective experimental details are given in the Supplementary Material.

\section*{Acknowledgments}

This paper is dedicated to Professor Klaus Betzler on the occasion of his 80th birthday, in recognition of his major contributions to this field. 

We thank the Group of Clara Saraceno for providing the Mg:sLN sample.
This work was funded by the Deutsche Forschungsgemeinschaft (DFG, German Research Foundation)—426703838 within the research unit ‘Periodic low-dimensional defect structures in polar oxides’-FOR 5044).

\section*{Author Contributions}

N.D. and T.H. conceived the idea. N.D. performed the optical experiments, analyzed the data and prepared the figures. T.H. and F.S. contributed to data analysis and interpretation. S.I. and S.G. provided the LNT crystals and performed the $\mu$-XRF measurements. M.I. was responsible for funding acquisition and project supervision. All authors discussed the results and contributed to the manuscript.

\subsection*{Competing interests}
The authors declare no competing interests.

\subsection*{Data Availability}

All data supporting the findings of this study are included in the paper and Supplementary Information. Additional raw data are available from the corresponding author upon reasonable request.

\bibliographystyle{naturemag}
\bibliography{bib}

\clearpage
\onecolumngrid

\setcounter{figure}{0}
\renewcommand{\thefigure}{S\arabic{figure}}
\renewcommand{\figurename}{Fig.}

\begin{center}
    {\Large \textbf{Supplementary Information}}\\[1em]
    {\large \textbf{Self-selected phase-matched second harmonic generation in nonlinear optical
materials: from phenomenon to applications}}\\[1em]
    Niklas Dömer, Tobias Hehemann, Felix Sauerwein, Sebastian Inckemann, Steffen Ganschow, and Mirco Imlau
\end{center}

\vspace{1em}

\setcounter{figure}{0}
\renewcommand{\thefigure}{S\arabic{figure}}

\begin{figure}[htbp]
    \centering
    \includegraphics[width=0.75\textwidth]{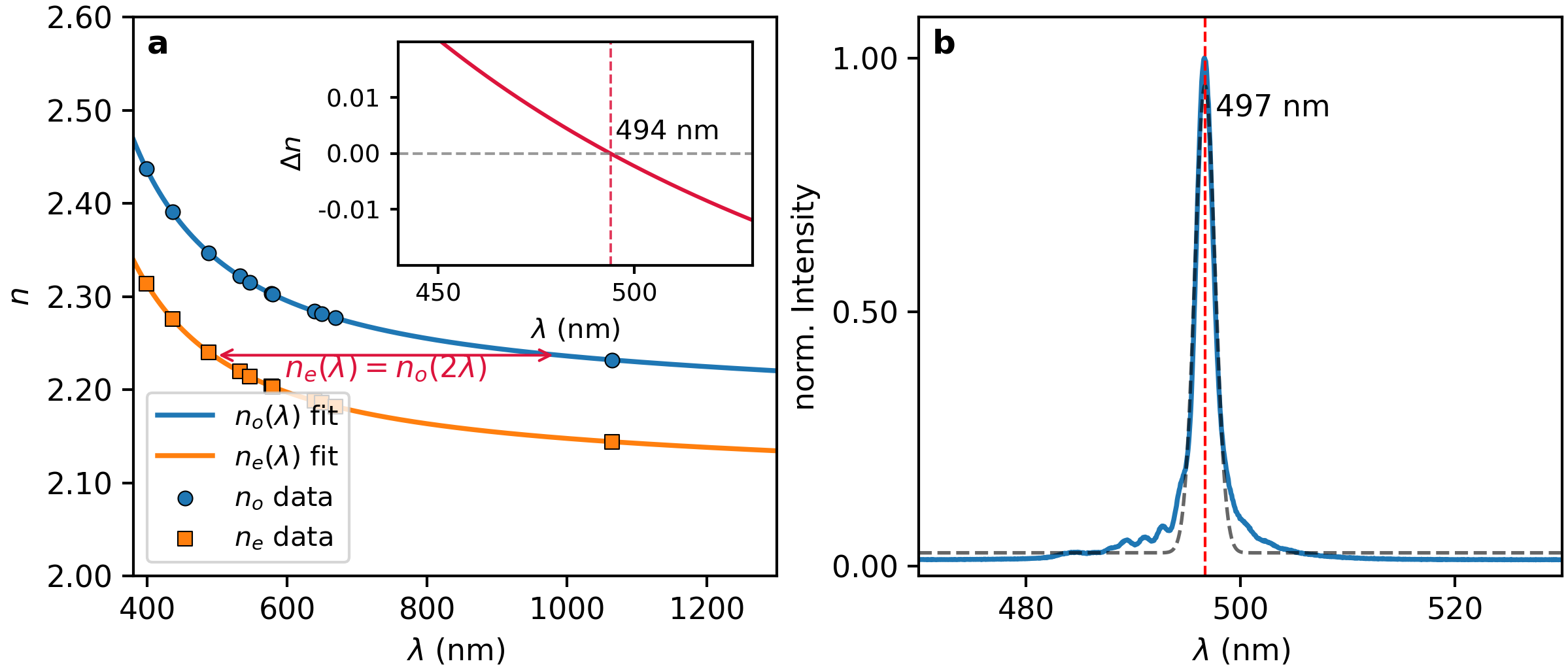}
    \caption{Measurement on 1.3 mol\% Mg-doped stoichiometric x-cut LN (Mg:sLN). The sample was supplied by Oxide Corporation and had a thickness of approximately $2\,\mathrm{mm}$. a: Ordinary and extraordinary refractive indices, $n_o$ and $n_e$, of Mg:SLN. Symbols show refractive-index data reported by Shoji \textit{et al.} \cite{Cleo_sln_Shoji:11}, while solid lines are fitted Sellmeier equations. The inset shows the corresponding noncritical phase-matching condition, $n_e(\lambda)=n_o(2\lambda)$, with $\lambda$ denoting the SHG wavelength. b: Measured SHG spectrum obtained with the OPA centered at $980\,\mathrm{nm}$. The dashed gray line shows a Gaussian fit to the SHG peak, and the red vertical line marks the fitted peak center. }
    \label{fig:Mgslm}
\end{figure}

\begin{figure}[htbp]
    \centering
    \includegraphics[width=0.75\textwidth]{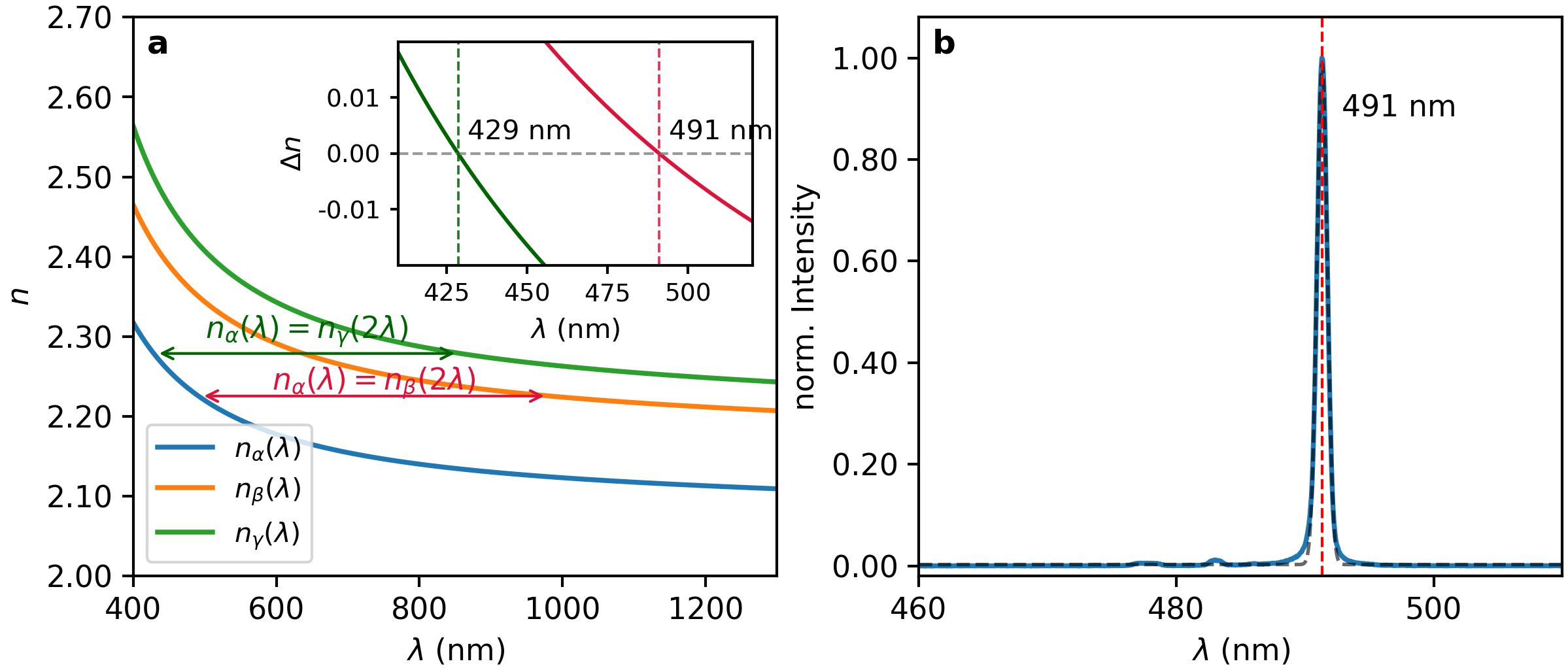}
    \caption{
Supplementary measurement on a $2.5\,\mathrm{mm}$-thick potassium niobate (KNbO$_3$) crystal supplied by SurfaceNet with light propagating along crystallographic $\gamma$ axis following the axis convention of Zysset \textit{et al.} \cite{KNB_refractive_index_Zysset:92}. 
a: Principal refractive indices $n_\alpha$, $n_\beta$, and $n_\gamma$ of orthorhombic KNbO$_3$ calculated from the 22~$^\circ$C Sellmeier coefficients reported by Zysset \textit{et al.} \cite{KNB_refractive_index_Zysset:92}. The inset shows the corresponding phase-matching conditions $n_\alpha(\lambda)=n_\beta(2\lambda)$ (red) and $n_\alpha(\lambda)=n_\gamma(2\lambda)$ (green), where $\lambda$ denotes the SHG wavelength. 
b: Measured SHG spectrum obtained under excitation at $970\,\mathrm{nm}$. The dashed gray line shows a Gaussian fit to the SHG peak, and the red vertical line marks the fitted peak center.
}
    \label{fig:knb}
\end{figure}

\begin{figure}[htbp]
    \centering
    \includegraphics[width=0.75\textwidth]{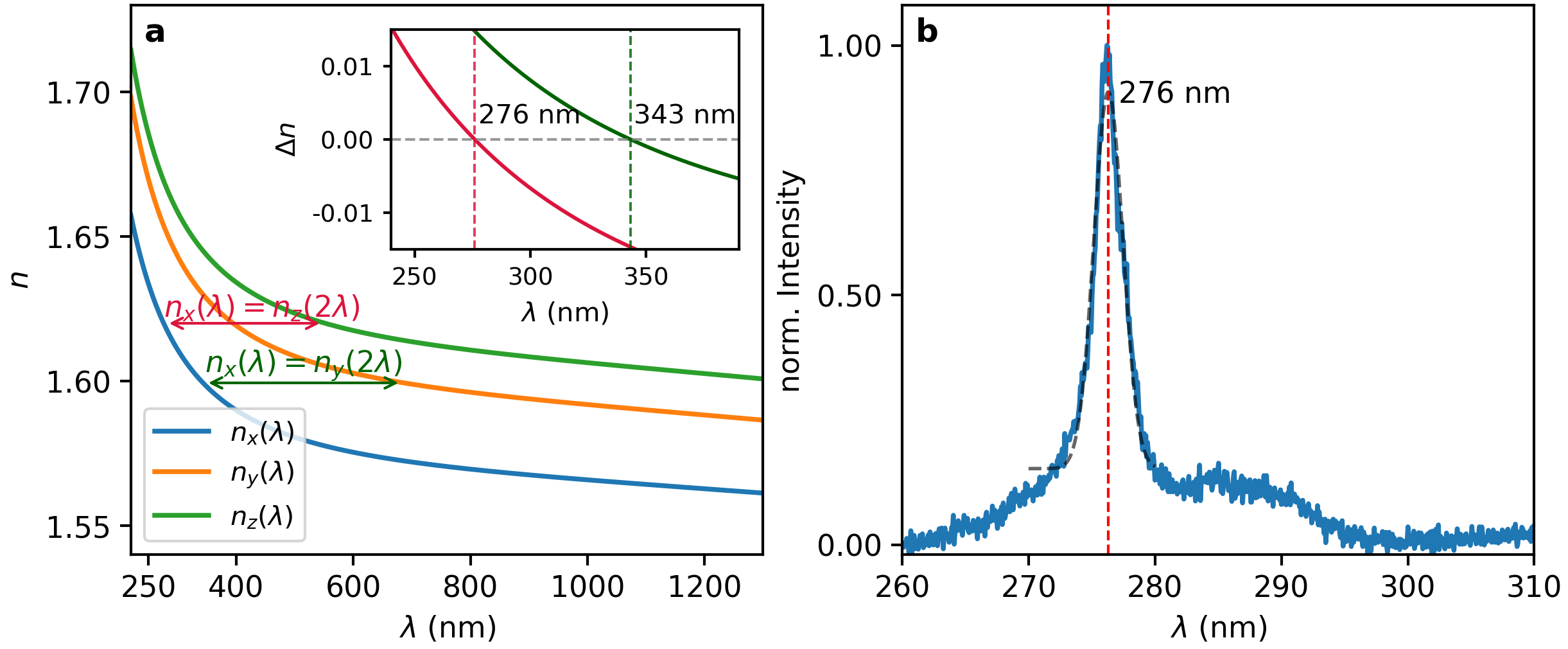}
    \caption{Supplementary measurement on a $3\,\mathrm{mm}$-thick $y$-cut lithium triborate (LBO) crystal. 
a: Principal refractive indices $n_x$, $n_y$, and $n_z$ of LBO calculated from Sellmeier equations \cite{LBO_refractive_index}. The inset shows the corresponding phase-matching conditions $n_x(\lambda)=n_z(2\lambda)$ (red) and $n_x(\lambda)=n_y(2\lambda)$ (green), where $\lambda$ denotes the SHG wavelength. 
b: Measured SHG spectrum obtained under excitation at $580\,\mathrm{nm}$. The dashed gray line shows a Gaussian fit to the SHG peak performed only in the range from $270$ to $280\,\mathrm{nm}$, and the red vertical line marks the fitted peak center.}
    \label{fig:LBO}
\end{figure}

\begin{figure}[h]
    \centering
    \includegraphics[width=0.75\textwidth]{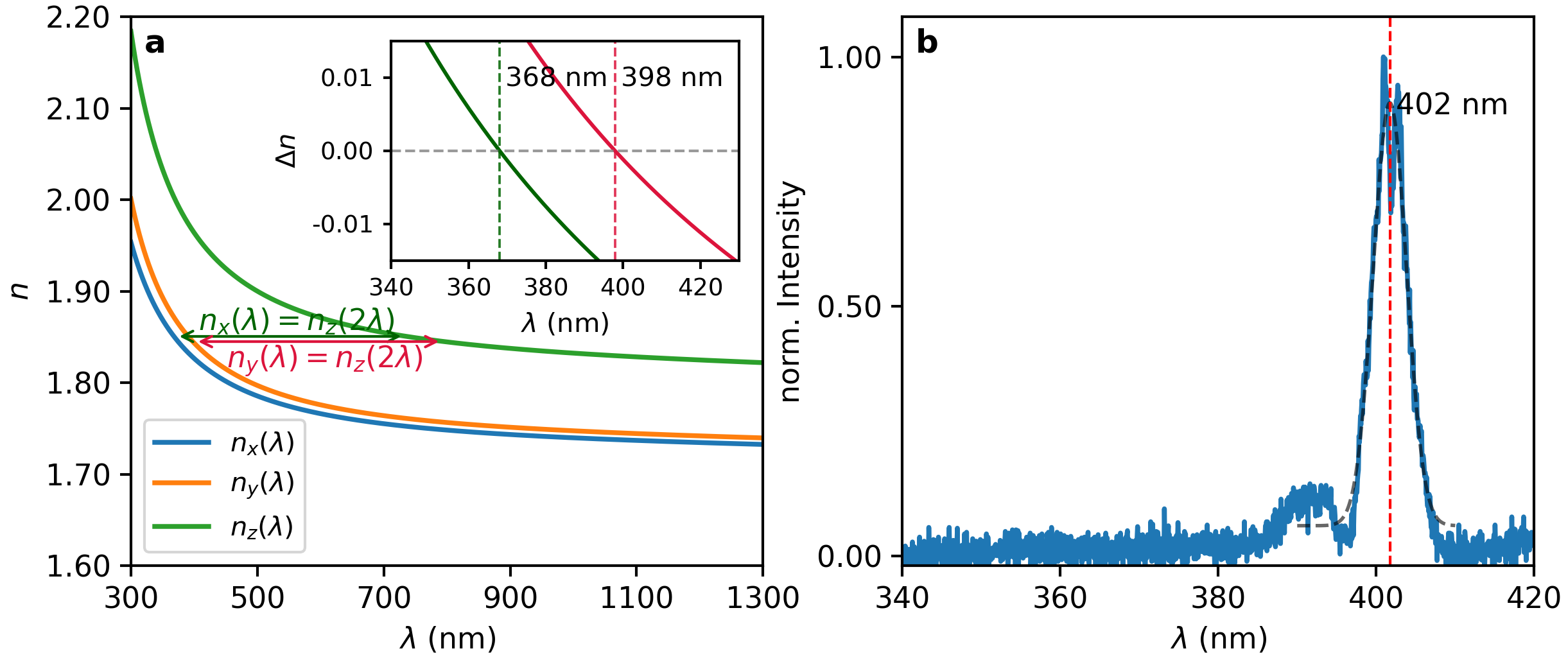}
    \caption{Supplementary measurement on a $1\,\mathrm{mm}$-thick x-cut KTP crystal supplied by Castech Inc. a: Principal refractive indices $n_x$, $n_y$, and $n_z$ of KTP calculated from Sellmeier equations \cite{KTP_Kato:02}. The inset shows the corresponding phase-matching conditions $n_x(\lambda)=n_z(2\lambda)$ (green) and $n_y(\lambda)=n_z(2\lambda)$ (red), where $\lambda$ denotes the SHG wavelength. b: Measured SHG spectrum obtained under excitation at $780\,\mathrm{nm}$. The dashed gray line shows a Gaussian fit performed only in the range from $390$ to $410\,\mathrm{nm}$, and the red vertical line marks the fitted peak center. A double-peak structure was reproducibly observed for several KTP crystals.}
    \label{fig:KTP}
\end{figure}

\end{document}